\documentclass[aps,pra,twocolumn,superscriptaddress,amssymb,showpacs]{revtex4}

\usepackage{graphicx}

\begin{document}

%\preprint{}

\title{Continuous variable entanglement of phase locked light beams}



\author{H.~H.~Adamyan}
\email[]{adam@unicad.am}
%\thanks{}
\affiliation{Yerevan State University, A. Manookyan 1, 375049,
Yerevan, Armenia} \affiliation{Institute for Physical Research,
National Academy of Sciences,\\Ashtarak-2, 378410, Armenia}

\author{G.~Yu.~Kryuchkyan}
\email[]{gkryuchk@server.physdep.r.am}
%\thanks{}
\affiliation{Yerevan State University, A. Manookyan 1, 375049,
Yerevan, Armenia} \affiliation{Institute for Physical Research,
National Academy of Sciences,\\Ashtarak-2, 378410, Armenia}


\begin{abstract}
We explore in detail the possibility of intracavity generation of
continuous-variable (CV) entangled states of light beams under
mode phase-locked conditions. We show that such quantum states can
be generated in self-phase locked nondegenerate optical parametric
oscillator (NOPO) based on a type-II phase-matched down-conversion
combined with linear mixer of two orthogonally polarized modes of
the subharmonics in a cavity. A quantum theory of this device,
recently realized in the experiment, is developed for both
sub-threshold and above-threshold operational regimes. We show
that the system providing high level phase coherence between two
generated modes, unlike to the ordinary NOPO, also exhibits
different types of quantum correlations between photon numbers and
phases of these modes. We quantify the CV entanglement as two-mode
squeezing and show that the maximal degree of the integral
two-mode squeezing(that is $50\%$ relative to the level of vacuum
fluctuations) is achieved at the pump field intensity close to the
generation threshold of self-phase locked NOPO, provided that the
constant of linear coupling between the two polarizations is much
less than the mode detunings. The peculiarities of CV entanglement
for the case of unitary, non-dissipative dynamics of the system
under consideration is also cleared up.
\end{abstract}

%\keywords{}

\maketitle

\section{INTRODUCTION}

It is now believed that entanglement of quantum composite systems
with continuous degree of freedom is the basis of most
applications in the field of Quantum Information \cite{q}.
Interest in continuous-variable (CV) entanglement is being
extensively excited by successful experiments on quantum
teleportation based on two-mode squeezed states \cite {levon2} as
well as the experiments dealing with entanglement in atomic
ensembles \cite {atom1}. Since then, remarkable theoretical and
experimental efforts have been devoted to generating and
quantifying CV entangled states.

In this paper we propose a novel type of CV entangled states of
light-field with reduced phase noise. They are different from the
well-known entangled Einstein-Podolsky-Rosen (EPR) states
generated in a nondegenerate optical amplifier
\cite{levon4,levon5}, which exhibit large phase fluctuations. We
believe that such entangled states of light-field with localized
phases can be generated in a self-phase locked nondegenerate
optical parametric oscillator (NOPO), based on the type-II
phase-matched down-conversion and additional phase locking process
stipulated by the intracavity waveplate. The motivations for this
study are the following:

For the first time the CV entangled states of light were studied
in \cite {levon4} and demonstrated experimentally in \cite{levon5}
for nondegenerate optical parametric amplifier (NOPA). Then a CV
entanglement source was built from two single-mode squeezed vacuum
states combined on a beamsplitter \cite {levon2}. It is well known
that each of the orthogonally polarized and frequency degenerate
fields generated by NOPO is a field of zero-mean values. The phase
sum of generated modes is fixed by the phase of the pump laser,
while their phase difference undergoes a phase diffusion process
\cite {levon4} stipulated by vacuum fluctuations. As a rule, the
NOPO phase diffusion noise is substantially greater than the shot
noise level, that limits the usage of NOPO in precision
phase-sensitive measurements. Various methods based on phase
locking mechanisms \cite{mason,fabr,murad,zond,kry} have been
proposed for reducing such phase diffusion. In the comparatively
simple scheme realized in the experiment \cite{mason}, self-phase
locking was achieved in NOPO by adding an intracavity quarter-wave
plate to provide polarization mixing between two orthogonally
polarized modes of the subharmonics. The evidence of self-phase
locking was provided there by the high level of phase coherence
between the signal and idler fields. Following this experiment,
the semiclassical theory of such NOPO was developed in
\cite{fabr}. Recently, the schemes of multiphoton parametric
oscillators based on cascaded down-conversion processes in $\chi
^{(2)}$ media placed inside the same cavity and showing self-phase
locking have been proposed \cite{murad}. As was demonstrated in
\cite{zond}, the system based on combination of OPO and second
harmonic generation also displays self-phase locking. The
formation of self-phase locking and its connection with squeezing
in the parametric four-wave mixing under two laser fields has been
demonstrated in \cite{kry}. An important characteristic of
self-phase locked devices concerns the phase structure of
generated subharmonics. Indeed, the formation of the variety of
distinct phase states under self-phase locked conditions has been
obtained in the mentioned Refs. \cite{mason,fabr,murad,zond,kry}.
It was recently noted that the schemes involving phase locking are
potentially useful for precise interferometric measurements and
optical frequency division because they combine fine tuning
capability and stability of type - II phase matching with
effective suppression of phase noise. That is why we believe it
will be interesting to consider phase locked dynamics also from
the perspective of Quantum Optics and, in particular, from the
standpoint of production of CV entanglement.

A further motivation for such task is connected with the problem
of experimental generation of bright entangled light. So far, to
the best of our knowledge, there is no experimental demonstration
of CV entanglement above the threshold of NOPO. The progress in
experimental study of bright two-mode entangled state from cw
nondegenerate optical parametric amplifier has been made in
\cite{levon8}. The theoretical investigation of CV entangled light
in transition through the generation threshold of NOPO is given in
\cite{new}. One of the principal experimental difficulties in
advance toward a high-intensity level is the impossibility to
control the frequency degeneration of modes above the threshold.
We hope that the usage of phase locked NOPO may open a new
interesting possibility to avoid this difficulty.

In this paper we report what is believed to be the first
investigation of self-phase locked CV entangled states. We develop
the quantum theory of self-phase locked NOPO, with decoherence
included, in application to the generation of such entangled
states. This scheme is based on the combination of two processes,
namely, type-II parametric down-conversion and linear polarization
mixing with cavity-induced feedback. The parametric
down-conversion is a standard technique used to produce an
entangled photon pairs as well as CV two-mode squeezed states
\cite{levon2}. The beam splitter including polarization mixer are
also considered as experimentally accessible devices for
production of entangled light-fields \cite{tan}. Besides these,
there have been some studies of a beam splitter for various
nonclassical input states, including two-mode squeezing states
\cite{kim}. It is obvious, and also follows from the results of
the mentioned papers \cite {murad,zond,kry}, that the operational
regimes of the combined system with cavity-induced feedback and
dissipation drastically differ from those for pure processes. We
show below that analogous situation takes place in the
investigation of quantum-statistical properties of a combined
system such as the self-phase locked NOPO.

The paper is arranged as follows. In Section II we formulate the
model of combined NOPO based on the processes of two-photon
splitting and polarization mixing, and present a semiclassical
analysis of the system. Section III is devoted to the analysis of
quantum fluctuations of both modes within the framework of
linearization procedure around the stable steady-state. In Section
IV we investigate the CV entangling resources of self-phase locked
NOPO on the base of two-mode squeezing for both sub-threshold and
above-threshold operational regimes. We also discuss in Section IV
the case of unitary, non-dissipative dynamics using a well
justified small-interaction time approximation. We summarize our
results in Section V.

\section{MODEL OF SELF-PHASE LOCKED NOPO}

\begin{figure}
\includegraphics[angle=-90,width=0.48\textwidth]{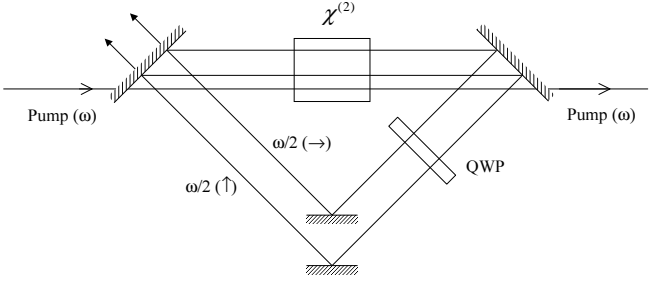}
\caption{The principal scheme of self-phase locked NOPO in triply
resonant cavity for the pump mode at frequency $\omega$ and two
subharmonic modes of two orthogonal polarizations ($\uparrow$) and
($\rightarrow$) at frequency $\omega/2$. We have chosen the
special path for the pump mode to underline that the pump mode
rapidly decays ($\gamma _{3}\gg \gamma_{1}, \gamma_{2}$) and is
eliminated adiabatically. The type-II phase-matching condition for
the process
$\omega\rightarrow\frac{\omega}{2}\left(\uparrow\right)+\frac{\omega}{2}\left(\rightarrow\right)$
is satisfied in $\chi^{\left(2\right)}$ medium. The intracavity
quarter waveplate QWP provides a polarization mixing between the
orthogonally polarized subharmonics.} \label{Scheme}
\end{figure}

As an entangler we consider the combination of two processes in a
triply resonant cavity, namely, the type - II parametric
down-conversion in $\chi ^{(2)}$- medium and polarization mixing
between subharmonics in lossless symmetric quarter-wave plate. The
Hamiltonian describing intracavity interactions is
\begin{eqnarray}
H &=&i\hbar E\left( e^{i\left( \Phi _{L}-\omega
t\right)}a_{3}^{+}-e^{-i\left( \Phi _{L}-\omega t\right)
}a_{3}\right)\nonumber \\
&&+i\hbar k\left(e^{i\Phi _{k}}a_{3}a_{1}^{+}a_{2}^{+}-e^{-i\Phi
_{k}}a_{3}^{+}a_{1}a_{2}\right)\nonumber \\
&&+\hbar \chi \left( e^{i\Phi _{\chi} }a_{1}^{+}a_{2}+e^{-i\Phi
_{\chi} }a_{1}a_{2}^{+}\right) , \label{OriginalH}
\end{eqnarray}
where $a_{i}$ are the boson operators for the cavity modes
$\omega_{i}$. The mode $a_{3}$ at frequency $\omega $ is driven by
an external field with amplitude $E$ and phase $\Phi _{L}$, while
$a_{1}$ and $a_{2}$ describe subharmonics of two orthogonal
polarizations at degenerate frequencies $\omega /2$. The constant
$ke^{i\Phi _{k}}$ determines the efficiency of the down-conversion
process. Linear coupling constant denoted as $\chi e^{i\Phi
_{\chi} }$describes the energy exchange between only the
subharmonic modes and besides, $\chi$ is determined by the amount
of polarization rotation due to the intracavity waveplate, $\Phi
_{\chi}$ determines the phase difference between the transformed
modes. We take into account the detunings of subharmonics $\Delta
_{i}$ and the cavity damping rates $\gamma _{i}$ and consider the
case of high cavity losses for pump mode ($\gamma _{3}\gg
\gamma_{1}, \gamma_{2}$), when this mode can be adiabatically
eliminated. However, in our analysis we take into account the pump
depletion effects. The principal scheme of this device, the
so-called self-phase locked NOPO, is shown on Fig.~\ref{Scheme}
for the special configuration when the coupling of in- and
out-radiation fields occurs at one of the ring cavity mirrors.
Note, that the comparison of this configuration with the analogous
one considered in \cite{fabr}, where the pump field is a
travelling wave, is discussed below.

Due to explicit presence of dissipation in this problem, one has
to write the master equation for the reduced density matrix of the
system. The reduced density operator $\rho $ within the framework
of the rotating wave approximation and in the interaction picture
is governed by the master equation
\begin{eqnarray}
\frac{\partial \rho }{\partial t}&=&\frac{1}{i\hbar }\left[H^{\prime},\rho %
\right]\nonumber\\
&+&\sum_{i=1}^{3}\gamma _{i}\left( 2a_{i}\rho
a_{i}^{+}-a_{i}^{+}a_{i}\rho -\rho a_{i}^{+}a_{i}\right),
\label{MasterEq}
\end{eqnarray}
where
\begin{eqnarray}
H^{\prime}&=&\sum_{i=1}^{3}\hbar \Delta _{i}a_{i}^{+}a_{i}+{i\hbar}E\left(a_{3}^{+}-a_{3}\right)\nonumber \\
&+&i\hbar k (a_{3}a_{1}^{+}a_{2}^{+}-a_{3}^{+}a_{1}a_{2}) +\hbar
\chi(a_{1}^{+}a_{2}+a_{1}a_{2}^{+}). \label{TransformedH}
\end{eqnarray}
Let us also note that this equation is rewritten through the
transformed boson operators $a_{i}\rightarrow a_{i}\exp \left(
-i\Phi _{i}\right) $, with $\Phi _{i}$ being $\Phi _{3}=\Phi
_{L}$, $\Phi _{2}=\frac{1}{2}\left(\Phi _{L}+\Phi _{k}-\Phi _{\chi
}\right)$, $\Phi _{1}=\frac{1}{2}\left(\Phi _{L}+\Phi _{k}+\Phi
_{\chi }\right)$. That leads to cancellation of the phases on the
intermediate stages of calculations. As a result, the Hamiltonian
$H^{\prime}$ depends only on real-valued coupling constants.

In order to proceed further, we now consider the phase-space
symmetry properties of the two subharmonic modes. It is easy to
check that the interaction Hamiltonian satisfies the commutation
relation $\left[ H^{\prime},U\left(\pi \right)\right] =0$ with the
operator $U\left(\theta \right)=\exp \left[ i\theta \left(
a_{1}^{+}a_{1}-a_{2}^{+}a_{2}\right) \right] $. Moreover,
analogous symmetry $\left[ \rho \left( t\right) ,U\left( \pi
\right) \right] =0$ takes place for the density operator of the
system, which obeys the master equation (\ref{MasterEq}). Using
this symmetry we establish the following selection rules for the
normal-ordered moments of the mode operators of the subharmonics:
\begin{equation}
\left\langle a_{1}^{+k}a_{2}^{+m}a_{1}^{l}a_{2}^{n}\right\rangle
=0, \label{ZeroMoments}
\end{equation}
if $k+l+m+n\neq 2j$, $j=0,\pm 1,\pm 2,...$. Another peculiarity
of\ NOPO with Hamiltonian (\ref{OriginalH}) is displayed in the
phase space of each of the subharmonic modes. The reduced density
operator for each of the modes is constructed from the density
operator $\rho $ by tracing over the other modes
$\rho_{1}=Tr_{2}Tr_{3}\left(\rho \right)$,
$\rho_{2}=Tr_{1}Tr_{3}\left(\rho \right)$. Therefore, we find that
$\left[ \rho _{1}\left( t\right) ,U_{1}\left( \pi \right) \right]
=\left[ \rho _{2}\left( t\right) ,U_{2}\left( \pi \right) \right]
=0$, where $U_{i}\left(\theta \right)=\exp \left(i\theta
a_{i}^{+}a_{i}\right),\;\left(i=1,2\right)$ is the rotation
operator. It is easy to check using these equations that the
Wigner functions $W_{1}$ and $W_{2}$ of the modes have a two-fold
symmetry under the rotation of the phase-space by angle $\pi $
around its origin,
\begin{equation}
W_{i}\left( r,\theta \right) =W_{i}\left( r,\theta +\pi \right) ,
\label{WignerSymmetry}
\end{equation}
where $r$, $\theta $ are the polar coordinates of the complex
phase space. Note, that such symmetry relationships
(\ref{ZeroMoments}), (\ref {WignerSymmetry}) radically differ from
those taking place for the usual NOPO without the quarter-wave
plate. Indeed, in this case, (case of $\chi =0$ in Eqs.
(\ref{OriginalH})-(\ref{TransformedH})) the Wigner functions
$W_{i}$ are rotationally symmetric and the symmetry properties
(\ref{ZeroMoments}) read as $\left\langle
a_{1}^{+k}a_{1}^{l}a_{2}^{+m}a_{2}^{n}\right\rangle =0$, if
$k-l\neq m-n$.

We perform concrete calculations in the positive P-representation
\cite{drum} in the frame of stochastic equations for the complex
c-number variables $\alpha _{i}$ and $\beta _{i}$ corresponding to
the operators $a_{i}$ and $a_{i}^{+}$
\begin{eqnarray}
\frac{\partial \alpha _{1}}{\partial t}&=&-\left(\gamma
_{1}+i\Delta_{1}\right)\alpha_{1}+k\alpha _{3}\beta _{2}-i\chi
\alpha _{2}+R_{1},\label{a1StochOriginalEq}\\
\frac{\partial \beta _{1}}{\partial t}&=&-\left( \gamma
_{1}-i\Delta _{1}\right) \beta _{1}+k\beta _{3}\alpha _{2}+i\chi
\beta _{2}+R_{1}^{+}, \label{b1StochOriginalEq}
\end{eqnarray}
\begin{eqnarray}
\frac{\partial \alpha _{3}}{\partial t}&=&-\gamma_{3}\alpha _{3} +
E - k\alpha_{1}\alpha_{2},\label{a3StochEq}\\
\frac{\partial \beta _{3}}{\partial t}&=&-\gamma_{3}\beta _{3} + E
- k\beta_{1}\beta_{2}.\label{b3StochEq}
\end{eqnarray}

The equations for $\alpha _{2}$, $\beta _{2}$ are obtained from
(\ref{a1StochOriginalEq} ), (\ref{b1StochOriginalEq}) by
exchanging the subscripts (1)$\rightleftarrows$(2). $R_{1,2}$ are
Gaussian noise terms with zero means and the following nonzero
correlators:
\begin{equation}
\left\langle R_{1}(t)R_{2}(t^{\prime })\right\rangle = k
\alpha_{3} \delta \left( t-t^{\prime }\right) ,
\label{R1R2CorrelatorOriginal}
\end{equation}
\begin{equation}
\left\langle R_{1}^{+}(t)R_{2}^{+}(t^{\prime })\right\rangle
=k\beta _{3}\delta \left( t-t^{\prime }\right) .
\label{R1PlusR2PlusCorrelatorOriginal}
\end{equation}
Recall, that we consider the regime of adiabatic elimination of
the pump mode. In this approach the stochastic amplitude $\alpha
_{3}$ and $\beta_{3}$ are given by
\begin{equation}
\alpha _{3}=\left( E-k\alpha_{1}\alpha _{2}\right) /\gamma
_{3}.\label{alpha3}
\end{equation}
\begin{equation}
\beta _{3}=\left( E-k\beta_{1}\beta _{2}\right) /\gamma
_{3}.\label{beta3}
\end{equation}
Substituting amplitudes $\alpha _{3}$, $\beta _{3}$ into
Eqs.(\ref{a1StochOriginalEq}), (\ref{b1StochOriginalEq}) and
(\ref{R1R2CorrelatorOriginal}),
(\ref{R1PlusR2PlusCorrelatorOriginal}) we arrive at
\begin{eqnarray}
\frac{\partial \alpha _{1}}{\partial t}&=&-\left(\gamma
_{1}+i\Delta
_{1}\right)\alpha_{1}\nonumber \\
&&+\left(\varepsilon -\lambda \alpha _{1}\alpha _{2}\right)\beta
_{2}-i\chi \alpha _{2}+R_{1}, \label{a1StochEq}
\end{eqnarray}
\begin{eqnarray}
\frac{\partial \beta _{1}}{\partial t}&=&-\left( \gamma
_{1}-i\Delta _{1}\right) \beta _{1}\nonumber \\
&&+\left( \varepsilon -\lambda \beta _{1}\beta _{2}\right) \alpha
_{2}+i\chi \beta _{2}+R_{1}^{+}. \label{b1StochEq}
\end{eqnarray}
\begin{equation}
\left\langle R_{1}(t)R_{2}(t^{\prime })\right\rangle =\left(
\varepsilon -\lambda \alpha _{1}\alpha _{2}\right) \delta \left(
t-t^{\prime }\right) , \label{R1R2Correlator}
\end{equation}
\begin{equation}
\left\langle R_{1}^{+}(t)R_{2}^{+}(t^{\prime })\right\rangle
=\left( \varepsilon -\lambda \beta _{1}\beta _{2}\right) \delta
\left( t-t^{\prime }\right) .  \label{R1PlusR2PlusCorrelator}
\end{equation}
Here $\varepsilon =kE/\gamma _{3}$, $\lambda =k^{2}/\gamma _{3}$.
So the Eqs.(\ref{a1StochEq}), (\ref{b1StochEq}) and corresponding
Eqs. for $\alpha _{2}$, $\beta _{2}$ involve the depletion effect
of the pump mode, which leads to the appearance of the
above-threshold operational regime.

It should be noted that Eqs.(\ref{a3StochEq}), (\ref{b3StochEq})
for pump mode amplitude do not involve terms of the linear
coupling between the modes of subharmonics. From this we can
immediately recognize that the Eqs.(\ref{alpha3}), (\ref{beta3})
and hence Eqs.(\ref{a1StochEq}), (\ref{b1StochEq}) take place for
arbitrary values of the parameter $\chi$. Nevertheless, the effect
of linear coupling between the modes are displayed in the dynamics
of pump mode through the amplitudes $\alpha_{1}$ and $\alpha_{2}$.
In the adiabatic approach these effects are described through the
pump depletion terms $k\alpha _{1}\alpha _{2}/\gamma _{3}$,
 $k\beta_{1}\beta _{2}/\gamma _{3}$ in the expressions for the
pump field amplitudes (\ref{alpha3}), (\ref{beta3}) and hence are
completely taken into consideration in Eqs.(\ref{a1StochEq}),
(\ref{b1StochEq}) as well as in the correlators
(\ref{R1R2Correlator}), (\ref{R1PlusR2PlusCorrelator}).

First, we shall study the steady-state solution of the stochastic
equations in semiclassical treatment, ignoring the noise terms for
the mean photon numbers $n_{j0}$ and phases $\varphi _{j0}$ of the
modes ($n_{j}=\alpha _{j}\beta _{j}$, $\varphi_{j}=\frac{1}{2i}\ln
\left(\alpha _{j}/\beta _{j}\right)$). The mean photon numbers
read as
\begin{equation}
n_{10}^{\pm }=\frac{1}{\lambda }\left( \frac{\Delta _{2}}{\Delta _{1}}%
\right) ^{1/2}\left[ \sqrt{\varepsilon ^{2}-\left( \varepsilon
_{cr}^{\pm }\right) ^{2}+\widetilde{\gamma
}^{2}}-\widetilde{\gamma }\right] , \label{Phot1ClassicGeneral}
\end{equation}
\begin{equation}
n_{20}^{\pm }=\frac{1}{\lambda }\left( \frac{\Delta _{1}}{\Delta
_{2}}\right) ^{1/2}\left[
\sqrt{\varepsilon ^{2}-\left( \varepsilon _{cr}^{\pm }\right) ^{2}+%
\widetilde{\gamma }^{2}}-\widetilde{\gamma }\right] ,
\label{Phot2ClassicGeneral}
\end{equation}
where
\begin{equation}
\widetilde{\gamma }=\frac{\gamma _{1}}{2}\left( \frac{\Delta
_{2}}{\Delta
_{1}}\right) ^{1/2}+_{{}}\frac{\gamma _{2}}{2}\left( \frac{\Delta _{1}}{%
\Delta _{2}}\right) ^{1/2}.  \label{GammaTilda}
\end{equation}
As we can see from (\ref{Phot1ClassicGeneral}), (\ref
{Phot2ClassicGeneral}), $n_{10}^{\pm }$ and $n_{20}^{\pm }$ are
real and positive for $\varepsilon $ exceeding two critical points
\begin{eqnarray}
\left(\varepsilon _{cr}^{\pm}\right)^{2}&=&\gamma _{1}\gamma
_{2}+\Delta _{1}\Delta
_{2}+\chi ^{2}\nonumber\\
&&\mp \sqrt{4\chi ^{2}\Delta _{1}\Delta _{2}-\left( \gamma
_{1}\Delta _{2}-\gamma _{2}\Delta _{1}\right) ^{2}},
\label{EpsilonCritical}
\end{eqnarray}
and, besides, the solutions $\ n_{i0}^{+}$ and $n_{i0}^{-},$
(i=1,2) correspond to two distinct values $\varepsilon _{cr}^{+}$
and $\varepsilon _{cr}^{-}$ accordingly. The steady-state values
of the phases corresponding to each of the critical points
$\varepsilon _{cr}^{-},$ $\varepsilon _{cr}^{+}$ are obtained as
\begin{eqnarray}
\sin ( \varphi _{20}^{+}&-&\varphi _{10}^{+} )
=\sin( \varphi_{20}^{-}-\varphi_{10}^{-} )=\nonumber\\
&=&\frac{1}{2\chi }\left[ \gamma _{1}\left( \frac{\Delta _{2}}{%
\Delta _{1}}\right) ^{1/2}-\gamma _{2}\left( \frac{\Delta _{1}}{\Delta _{2}}%
\right) ^{1/2}\right] ,  \label{SinPhaseDiff}
\end{eqnarray}
\begin{equation}
\cos (\varphi _{20}^{\pm }+\varphi _{10}^{\pm })=\frac{1}{\varepsilon }\sqrt{%
\varepsilon ^{2}-\left( \varepsilon _{cr}^{\pm }\right) ^{2}+\widetilde{%
\gamma }^{2}}.  \label{CosPhaseSum}
\end{equation}
\ It is easy to check that these solutions exist for both modes
only if the following relation holds
\begin{equation}
4\chi ^{2}\Delta _{1}\Delta _{2}>\left( \gamma _{1}\Delta
_{2}-\gamma _{2}\Delta _{1}\right) ^{2}.
\label{SolutionCondition}
\end{equation}
Let us note that the steady-state solutions (\ref{SinPhaseDiff}),
(\ref {CosPhaseSum}) completely determine the absolute phases of
the orthogonally polarized modes, which are hence self-locked,
unlike the ordinary NOPO. These results are in accordance with the
ones obtained in \cite {mason,fabr}, but for another configuration
of NOPO. In the scheme proposed in \cite{mason} only the signal
and idler modes are excited in the cavity, while the pump field is
a travelling wave. Nevertheless, in the adiabatic regime
considered here there is correspondence between both schemes.
Indeed, it is not difficult to check that the results (\ref
{Phot1ClassicGeneral}), (\ref {Phot2ClassicGeneral}) transform to
the corresponding results of the mentioned scheme \cite {fabr} by
replacing the parameter $\varepsilon $ with the corresponding pump
field amplitude.

We now turn to the standard linear stability analysis of these
solutions, assuming for simplicity, the perfect symmetry between
the modes, provided that the cavity decay rates and the detunings
do not depend on the polarization ( $\gamma _{1}=\gamma
_{2}=\gamma $, $\Delta _{1}=\Delta _{2}=\Delta $). The stability
of the system is governed by the matrices $F$ and $F_{+},$ $F_{-}$
describing the dynamics of small deviations $\delta \alpha
_{i\text{ }}$and $\delta \beta _{i}$ from the semiclassical
steady-state solutions (see, Sec.III). We reach stability if the
real parts of eigenvalues of these matrices are positive. This
analysis displays an evident dependence on the sign of the
detunings $\Delta _{1}=\Delta _{2}=\Delta $. For the positive
detuning, $\Delta >0$, only the steady-state solutions $%
n_{10}^{-}=n_{20}^{-} $ and $\varphi _{20}^{-}-\varphi _{10}^{-}$,
$\varphi _{20}^{-}+\varphi _{10}^{-}$ are stable, while for the
case of negative detuning the stability holds for the solutions
with the (+) superscript. As this analysis shows, for either sign
of detuning, the threshold is reached at $\varepsilon \geqslant $
$\varepsilon _{th}$, where
\begin{equation}
\varepsilon _{th}=\sqrt{\left( \chi -\left| \Delta \right| \right)
^{2}+\gamma ^{2}},  \label{EpsilonThreshold}
\end{equation}
and the steady-state stable solution for mean photon numbers can
be written in the general form as
\begin{equation}
n_{0}=n_{10}=n_{20}=\frac{1}{\lambda }\left[ \sqrt{\varepsilon
^{2}-\left( \chi -\left| \Delta \right| \right) ^{2}}-\gamma
\right] . \label{Phot12Simplified}
\end{equation}
The phases are found to be
\begin{equation}
\varphi _{10}=\varphi _{20}=-\frac{1}{2}Arc\sin
\frac{1}{\varepsilon }\left( \chi +\left| \Delta \right| \right)
+\pi k,  \label{PhasesNegDelta}
\end{equation}
for $\Delta >0$. For the opposite sign of the detuning, $\Delta
<0$, as we noted, the mean photon numbers are given by the same
Eq.(\ref {Phot12Simplified}), while the phases read as
\begin{eqnarray}
\varphi _{10} &=&\frac{1}{2}Arc\sin \frac{1}{\varepsilon }\left(
\chi +\left| \Delta \right| \right) +\pi \left(
k+\frac{1}{2}\right) ,
\label{PhasesPosDelta} \\
\varphi _{20} &=&\frac{1}{2}Arc\sin \frac{1}{\varepsilon }\left(
\chi +\left| \Delta \right| \right) +\pi \left(
k-\frac{1}{2}\right) ,  \nonumber
\end{eqnarray}
($k=0,1,2,..$). In the region $\varepsilon \leqslant $
$\varepsilon _{th}$ the stability condition is fulfilled only for
the zero amplitude steady-state solution $\alpha _{1}=\alpha
_{2}=\beta _{1}=\beta _{2}=0$. So, the set of above-threshold
stable solutions for both modes have two-fold symmetry in the
phase-space which was indeed expected from symmetry arguments
(\ref{WignerSymmetry}).

Let us now consider the output behavior of the system for the
special scheme of generation, when the couplings of in- and
out-fields occur at only one of the ring-cavity mirrors (see
Fig.~\ref{Scheme}). Taking into account that only the fundamental
mode is coherently driven by the external field with $\langle
\alpha _{3}^{in}\rangle =E/\gamma _{3}$, while the subharmonic
modes are initially in the vacuum state, we obtain for the mean
photon numbers (in units of photon number per unit time)
$n_{3}^{in}=E^{2}/2\gamma _{3},\;n_{i}^{out}=2\gamma
_{i}n_{i0}\;\left(i=1,2\right)$ and hence
$n_{1}^{out}=n_{2}^{out}=2\gamma n_{0}$. Accordingly, parametric
oscillation can occur above the threshold pump power $P_{th}=\hbar
\omega ^{3} E_{th}^{2}/2\gamma _{3}$, where the threshold value of
the pump field is equal to
\begin{equation}
E_{th}=\frac{\gamma _{3}}{k}\sqrt{\left(\chi -\left| \Delta
\right| \right)^{2}+\gamma ^{2}}. \label{EThreshold}
\end{equation}

We are now in a position to study quantum effects in self-phase
locked NOPO and will state the main results of the paper
concerning CV entanglement.

\section{ANALYSIS OF QUANTUM FLUCTUATIONS}

The aim of the present section is to study the quantum-statistical
properties of self-phase locked NOPO in linear treatment of
quantum fluctuations. Quantum analysis of the system using $P$-
representation is standard. A detailed description of the method
can be found in \cite{drum}. We assume that the quantum
fluctuations are sufficiently small so that Eqs.(\ref{a1StochEq}),
(\ref{b1StochEq}) can be linearized around the stable
semiclassical steady state $\alpha _{i}\left( t\right) =\alpha
_{i}^{0}+\delta \alpha _{i}\left( t\right) $, $\beta _{i}\left(
t\right) =\beta _{i}^{0}+\delta \beta _{i}\left( t\right) $. This
is appropriate for analyzing the quantum-statistical effects,
namely CV entanglement, for all operational regimes with the
exception of the vicinity of threshold, where the level of quantum
noise increases substantially. It should also be emphasized at
this stage that in the above-threshold regime of self-phase locked
NOPO the steady-state phases of each of the modes are well-defined
in contrast to what happens in the case of ordinary NOPO, where
phase diffusion takes place. According to this effect, the
difference between the phases, as well as each of the phases, can
not be defined in the above-threshold regime of generation of the
ordinary NOPO. On the whole, the well founded linearization
procedure can not be applied for this case. Nevertheless, the
linearization procedure and analysis of quantum fluctuations for
ordinary NOPO become possible due to the additional assumptions
about temporal behavior of the difference between the phases of
the generated modes \cite{levon10}.

We begin with consideration of below-threshold operational regime,
for $E<E_{th}$, where the equations linearized around the
zero-amplitude solution can be written in the following matrix
form
\begin{equation}
\frac{\partial }{\partial t}\delta \alpha ^{\mu }=-F_{\mu \nu
}\delta \alpha ^{\nu }+R^{\mu }\left( \alpha ,t\right) ,
\label{daStochEq}
\end{equation}
where $\mu =1,2,3,4$ and $\delta \alpha ^{\mu }=\left( \delta
\alpha _{1},\delta \alpha _{2},\delta \alpha _{3},\delta \alpha
_{4}\right) =\left( \delta \alpha _{1},\delta \alpha _{2},\delta
\beta _{1},\delta \beta _{2}\right) $, $R^{\mu }=\left(
R_{1},R_{2},R_{1}^{+},R_{2}^{+}\right) $. The 4$\times $4 matrix
$F_{\mu \nu }$ is written in the block form
\begin{equation}
F=\left(
\begin{array}{cc}
A, & B \\
B^{\ast }, & A^{\ast }
\end{array}
\right)  \label{FMatrix}
\end{equation}
with 2$\times $2 matrices
\begin{equation}
A=\left(
\begin{array}{cc}
\gamma +i\delta, & i\chi \\
i\chi, & \gamma +i\delta
\end{array}
\right) ,\;\;B=\varepsilon \left(
\begin{array}{cc}
0, & 1 \\
1, & 0
\end{array}
\right) .  \label{AandBMatrices}
\end{equation}
The noise correlators are determined as
\begin{equation}
\left\langle R^{\mu }\left( \alpha ,t\right) R^{\nu }\left( \alpha
,t^{\prime }\right) \right\rangle =D_{\mu \nu }\left( \alpha
\right) \delta \left( t-t^{\prime }\right)  \label{R0Correlators}
\end{equation}
with the following diffusion matrix
\begin{equation}
D=\left(
\begin{array}{cc}
B, & 0 \\
0, & B^{\ast }
\end{array}
\right) .  \label{DMatrix}
\end{equation}

First, we calculate the temporal correlation functions of the
fluctuation operators. The expectation values of interest can be
written as the integral
\begin{equation}
\left\langle \delta \alpha ^{\mu }\left( t\right) \delta \alpha
^{\nu }\left( t^{\prime }\right) \right\rangle =\int_{-\infty
}^{\min \left( t,t^{\prime }\right) }d\tau \left( e^{F\left(
t-\tau \right) }De^{F^{T}\left( t^{\prime }-\tau \right) }\right)
_{\mu \nu }, \label{dadaCorrelator}
\end{equation}
$F^{T}$ being the transposition of the matrix $F$. The integration
over $d\tau$ can be performed using the following useful formula
for operators $DF^{T}=FD$, obtained by straightforward
calculation. As a consequence, we arrive at the expression
$De^{F^{T}t}=e^{Ft}D$. Finally we obtain
\begin{equation}
\left\langle \delta \alpha ^{\mu }\left( t\right) \delta \alpha
^{\nu }\left( t^{\prime }\right) \right\rangle =-\frac{1}{2}\left(
F^{-1}e^{F\left| t-t^{\prime }\right| }D\right) _{\mu \nu },
\label{dadaCorrelatorSimplified}
\end{equation}
and hence for $t=t^{\prime }$%
\begin{equation}
\left\langle \delta \alpha ^{\mu }\left( t\right) \delta \alpha
^{\nu }\left( t\right) \right\rangle =-\frac{1}{2}\left(
F^{-1}D\right) _{\mu \nu }.  \label{dadaMean}
\end{equation}

This formula, however, is not very convenient for practical
calculations. Therefore, we rewrite the correlation functions of the quantum fluctuations $%
\delta \alpha _{i}$, $\delta \beta _{i}$ in a more simple form
through the two-dimensional column vectors $\delta \alpha =\left(
\delta \alpha _{1},\delta \alpha _{2}\right) ^{T}$, $\delta \beta
=\left( \delta \beta _{1},\delta \beta _{2}\right) ^{T}$. Upon
performing the calculation, we finally arrive at
\begin{widetext}
\begin{eqnarray}
\left\langle \delta \alpha \left( \delta \alpha \right)
^{T}\right\rangle &=&\frac{\varepsilon }{S^{4}-4\Delta ^{2}\chi
^{2}}\left[ \gamma \left(
\begin{array}{cc}
-2\chi \Delta, & S^{2} \\
S^{2}, & -2\chi \Delta
\end{array}
\right) -i\left(
\begin{array}{cc}
\chi \left(S^{2} -  2\Delta ^{2}\right), & \Delta \left(S^{2} -  2\chi ^{2}\right) \\
\Delta \left(S^{2} -  2\chi ^{2}\right), & \chi \left(S^{2} -
2\Delta ^{2}\right)
\end{array}
\right) \right],  \nonumber \\
  \nonumber \\
\left\langle \delta \alpha \left( \delta \beta \right) ^{T
}\right\rangle &=&\frac{\varepsilon ^{2}}{2\left(S^{4}-4\Delta
^{2}\chi ^{2}\right)}\left(
\begin{array}{cc}
S^{2}, & -2\chi \Delta \\
-2\chi \Delta, & S^{2}
\end{array}
\right) , \label{dadbMeanFinal}
\end{eqnarray}
\end{widetext}
where $S^{2}$ is introduced as
\begin{equation}
S^{2}=\gamma ^{2}+\chi ^{2}+\Delta ^{2}-\varepsilon ^{2}.
\label{Determinant}
\end{equation}
Note, that $S^{2}>0$ in the below-threshold regime.

As an application of these results the mean photon number in the
below-threshold regime can be calculated as follows,
\begin{equation}
n_{1}=n_{2}=\frac{\varepsilon ^{2}S^{2} }{2\left(S^{4}-4\Delta
^{2}\chi ^{2}\right) }. \label{PhotQBelow}
\end{equation}

Next we focus on the mode locked regime, for $E>E_{th}$,
considering the Eqs. (\ref{a1StochEq}), (\ref{b1StochEq}) in terms
of the fluctuations $\delta n_{i}\left( t\right) =n_{i}\left(
t\right) -n_{i0}$ and $\delta \varphi _{i}\left( t\right) =\varphi
_{i}\left( t\right) -\varphi _{i0}$ of photon number and phase
variables. In this regime the dynamics described by the linearized
equations of motion actually decouples into two independent
dynamics for two groups of combinations $\delta n_{\pm }=\delta
n_{2}\pm \delta n_{1}$, $\delta \varphi _{\pm }=\delta \varphi
_{2}\pm \delta \varphi _{1}$. In fact, one has
\begin{equation}
\frac{\partial }{\partial t}%
{\delta n_{+} \choose \delta \varphi _{+}} =-F_{+} {\delta n_{+}
\choose \delta \varphi _{+}} + {R_{n+} \choose R_{\varphi +}},
\label{dfiPlus_dnPlusStochEq}
\end{equation}
\begin{equation}
\frac{\partial }{\partial t} {\delta n_{-} \choose \delta \varphi
_{-}}=-F_{-}{\delta n_{-} \choose \delta \varphi _{-}} +{R_{n_{-}}
\choose R_{\varphi _{-}}}.  \label{dfiMinus_dnMinusStochEq}
\end{equation}
The drift, $F_{\pm }$, and the diffusion matrices, $\left\langle
R_{i}\left( t\right) R_{j}\left( t^{\prime }\right) \right\rangle
=D_{ij}^{(+)}\delta
\left( t-t^{\prime }\right) $, $(i,j=n_{+},\delta \varphi _{+})$; $%
\left\langle R_{n}\left( t\right) R_{m}\left( t^{\prime }\right)
\right\rangle =D_{nm}^{(-)}\delta \left( t-t^{\prime }\right)
,(n,m=n_{-},\varphi _{-})$ are respectively
\begin{equation}
\ F_{+}=\left(
\begin{array}{cc}
2\lambda n_{0}, & 4n_{0}\varepsilon \sin (\varphi _{20}+\varphi _{10}) \\
0, & 2\left( \gamma +\lambda n_{0}\right)
\end{array}
\right) ,  \label{F+Matrix}
\end{equation}
\begin{equation}
F_{-}=\left(
\begin{array}{cc}
2\gamma , & 4n_{0}\chi \sin n/\Delta \\
\Delta /n_{0}, & 0
\end{array}
\right) ,  \label{F-Matrix}
\end{equation}
\begin{equation}
D^{\left( \pm\right) }=\pm\left(
\begin{array}{cc}
4n_{0}\gamma ,~ -2\varepsilon \sin (\varphi _{20}+\varphi _{10}) &  \\
-2\varepsilon \sin (\varphi _{20}+\varphi _{10}),~ -\gamma /n_{0}
&
\end{array}
\right) .  \label{D+-Matrix}
\end{equation}

Due to the decoupling between $\left( +\right) $ and $\left(
-\right) $ combinations of the modes, we conclude that the
following temporal correlation functions are equal to zero,
$\left\langle \delta \varphi _{+}\left(t\right)\delta \varphi
_{-}\left(t^{\prime }\right)\right\rangle =\left\langle \delta
n_{+}\left(t\right)\delta n_{-}\left(t^{\prime
}\right)\right\rangle=\left\langle \delta
n_{\pm}\left(t\right)\delta \varphi_{\mp}\left(t^{\prime
}\right)\right\rangle=0$, and hence $\left\langle \left( \delta
\varphi _{1}\right) ^{2}\right\rangle =\left\langle \left( \delta
\varphi _{2}\right) ^{2}\right\rangle $, $\left\langle \left(
\delta n_{1}\right) ^{2}\right\rangle =\left\langle \left( \delta
n_{2}\right) ^{2}\right\rangle $. The other correlation functions
can be calculated in the same way as described above for the
below-threshold regime. The temporal correlation functions are
derived as
\begin{equation}
\left\langle{\delta n_{\pm }\left( t\right)  \choose \delta
\varphi _{\pm }\left( t\right) }\left( \delta n_{\pm }\left(
t^{\prime }\right) ,\delta \varphi _{\pm }\left( t^{\prime
}\right) \right) \right\rangle =-\frac{1}{2}F_{\pm
}^{-1}e^{-F_{\pm }^{-1}\left| t-t^{\prime }\right| }D_{\pm },
\label{dn+-dfi+-Correlator}
\end{equation}
and hence
\begin{equation}
\left\langle{\delta n_{\pm } \choose \delta \varphi _{\pm }}\left(
\delta n_{\pm },\delta \varphi _{\pm }\right) \right\rangle
=-\frac{1}{2}F_{\pm }^{-1}D_{\pm }.  \label{dn+-dfi+-Mean}
\end{equation}

Performing the concrete calculations for each of the cases $\Delta >0$ and $%
\Delta <0$, we arrive at the following results
\begin{widetext}
\begin{eqnarray}
\left\langle{\delta n_{+} \choose \delta \varphi _{+}}\left(
\delta n_{+},\delta \varphi _{+}\right) \right\rangle
&=&\frac{1}{4\lambda n_{0}(\gamma +\lambda n_{0})}\left(
\begin{array}{cc}
4n_{0}\left[ \gamma (\gamma +\lambda n_{0})+(\chi -\mid \Delta
\mid )^{2}\right] , & -\lambda n_{0}(\chi -\mid \Delta \mid ) sign(\Delta ) \\
-2\lambda n_{0}(\chi -\mid \Delta \mid )sign(\Delta ), & -\lambda
\gamma
\end{array}
\right) ,  \label{dn+dfi+MeanFinal}\\
\nonumber\\
\left\langle {\delta n_{-} \choose \delta \varphi _{-}} \left(
\delta n_{-},\delta \varphi _{-}\right) \right\rangle
&=&\frac{1}{4\mid \Delta \mid \chi }\left(
\begin{array}{cc}
4n_{0}\chi (\chi -\mid \Delta \mid ), & 2\chi \gamma sign(\Delta ) \\
2\chi \gamma sign(\Delta ), & \frac{1}{n_{0}}(\gamma ^{2}-\mid
\Delta \mid (\chi -\mid \Delta \mid )
\end{array}
\right) .  \label{dn-dfi-MeanFinal}
\end{eqnarray}
\end{widetext}

We see that the considered system displays different types of
quantum correlations in terms of the stochastic variables, namely,
between photon-number sum and phase sum in the modes, as well as
between photon-number difference and phase difference in the
modes. These results radically differ from the ones taking place
for usual NOPO, where the correlation between photon-number
difference and phase sum in the modes is realized. The results
obtained indicate the possibilities to produce entanglement with
respect to the new types of quantum correlations.

\section{CV ENTANGLEMENT IN THE PRESENCE OF PHASE LOCKING}

Let us now turn our attention to quantum statistical effects and
the entanglement production for the case of perfect symmetry between the modes $%
(\gamma _{1}=\gamma _{2}=\gamma ,$\ $\Delta _{1}=\Delta
_{2}=\Delta )$. Our aim in this section is to study the interplay
between phase locking phenomena and CV entanglement for the
self-phase locked NOPO. We note that unlike the two-mode squeezed
vacuum state, the state generated in the above-threshold regime of
NOPO is non-Gaussian, i.e. its Wigner function is non-Gaussian
\cite{cirak}. Recently, it has been demonsrated \cite{simon}, that
some systems involving beam splitters also generate non-Gaussian
states. The general consideration of this problem for self-phase
locked NOPO seems to be very complicated. However, the mentioned
results allow us to conclude that the state generated in
self-phase locked NOPO is most probably non-Gaussian. So far, the
inseparability problem for bipartite non-Gaussian state is far
from being understood. On the theoretical side, the necessary and
sufficient conditions for the separability of bipartite CV systems
have been fully developed only for Gaussian states, which are
completely characterized by their first and second moments. To
characterize the CV entanglement we address to both the
inseparability and strong EPR entanglement criteria \cite{epr}
which could be quantified by analyzing the variances of the
relevant distance $V_{-}=V\left( X_{1}-X_{2}\right) $ and the
total momentum $V_{+}=V\left( Y_{1}+Y_{2}\right)$ of the
quadrature amplitudes of two modes $X_{k}=\frac{1}{\sqrt{2}}\left[
a_{k}^{+}\exp \left( -i\theta _{k}\right) + a_{k}\exp \left(
i\theta _{k}\right) \right]$, $Y_{k}=\frac{i}{\sqrt{2}}\left[
a_{k}^{+}\exp \left( -i\theta _{k}\right) - a_{k}\exp \left(
i\theta _{k}\right) \right]$, $\left(k=1,2\right)$, where
$V(X)=\left\langle X^{2}\right\rangle -\left\langle X\right\rangle
^{2}$ is a denotation for the variance and $\theta _{k}$ is the
phase of local oscillator for the k-th mode. The two quadratures
$X_{k}$ and $Y_{k}$ are non commuting observables. The
inseparability criterion for the quantum state of two optical
modes reads as \cite{epr}
\begin{equation}
V=\frac{1}{2}(V_{+}+V_{-})<1,  \label{EntanglementCriteria}
\end{equation}
i.e. indicates that the sum of variances drops below the level of
vacuum fluctuations. Since the states of the system considered are
non-Gaussian, the criterion (\ref{EntanglementCriteria}) is only
sufficient for inseparability. The strong EPR entanglement
criterion is quantified by the product of variances as
$V_{+}V_{-}<\frac{1}{4}$. We remind that the sufficient condition
for inseparability (\ref{EntanglementCriteria}) in terms of the
product of variances reads as $V_{+}V_{-}<1$, i.e. is weaker than
the strong EPR condition.

In order to obtain the general expressions for variances, we first
write them in terms of the boson operators corresponding to the
Hamiltonian (\ref {OriginalH}). We perform the transformations
$a_{i}\rightarrow a_{i}\exp \left( i\Phi _{i}\right)$, which
restore the previous phase structure of the intracavity
interaction. Using also the symmetry relationships (\ref
{ZeroMoments}) we find quite generally the variances at some
arbitrary quadrature phase angles $\theta _{1}$, $\theta _{2}$ as
\begin{equation}
V_{\pm }=V\pm R\cos (\Delta \theta), \label{V+-ByVandR}
\end{equation}
where
\begin{eqnarray}
V &=&\frac{1}{2}(V_{+}+V_{-})=1+2n  \nonumber \\
&-2\mid &\left\langle a_{1}a_{2}\right\rangle \mid \cos (\Sigma
\theta+\Phi _{\arg }),~ \label{Vby_a}
\end{eqnarray}
\begin{equation}
R=2\mathop{\rm Re}(\left\langle a_{1}^{2}\right\rangle e^{i\Sigma
\theta })-2\mid \left\langle a_{1}^{+}a_{2}\right\rangle \mid ,
\label{RBy_a}
\end{equation}
\begin{equation}
\Delta \theta = \theta _{2}-\theta _{1}-\Phi_{\chi}, ~\Sigma
\theta = \theta _{1}+\theta _{2}+\Phi _{l}+\Phi _{k},
\label{DeltaSigmaTetta}
\end{equation}
and $n=\left\langle a_{1}^{+}a_{1}\right\rangle =\left\langle
a_{2}^{+}a_{2}\right\rangle$ is the mean photon number of the
modes, $\Phi _{\arg }=\arg \left\langle a_{1}a_{2}\right\rangle $.
So, in accordance with the formula (\ref{V+-ByVandR}), the
relative phase $\Phi_{\chi} $ between the transformed modes gives
the effect of the rotation of the quadrature amplitudes angle
$\theta _{2}-\theta _{1}$.

Obviously, the variances $V_{\pm }$ and hence the level of CV
entanglement depend on all parameters of the system including the
phases. The minimal possible level of $V$ is realized for an
appropriate selection of the phases $\theta _{i}$, namely for
$\theta _{1}+\theta _{2}=-\arg \left\langle
a_{1}a_{2}\right\rangle -$ $\Phi _{l}-\Phi _{k}$. Further, in most
cases we assume that this phase relationship takes place, but do
not introduce new denotations for $V$ and $V_{\pm}$ for the sake
of simplicity. In this case, ($\Sigma\theta=-\Phi _{\arg }$), in
correspondence with the formula (\ref {V+-ByVandR}), the variances
$V_{\pm}$ depend only on the difference between phases $\Delta
\theta$ and we arrive at
\begin{equation}
V=1+2(n-\mid \left\langle a_{1}a_{2}\right\rangle \mid ),
\label{VminBy_a}
\end{equation}
\begin{equation}
R=2\mathop{\rm Re} (\left\langle a_{1}^{2}\right\rangle e^{-i\Phi
_{_{\arg }}})-2\mid \left\langle a_{1}^{+}a_{2}\right\rangle \mid
. \label{RBy_aForMin}
\end{equation}

For the NOPO without additional polarization mixing $\left\langle
a_{1}^{+}a_{2}\right\rangle =\left\langle a_{1}^{2}\right\rangle
=\left\langle a_{2}^{2}\right\rangle =0$ and hence the case of the
symmetric variances $V_{+}=V_{-}$ is realized. The phase-locked
NOPO generally has non symmetric uncertainty region. However, the
variances $V_{+}$ and $V_{-}$ become equal for the special case of
$\theta _{2}-\theta _{1}-\Phi_{\chi} =\frac{ \pi }{2}$, when the
inseparability condition reads as $V<1$, $\left(
V=V_{-}=V_{+}\right) $. We note, that the relative phase
$\Phi_{\chi}$ plays an important role in specification of the
entanglement.

\subsection{Entanglement in the case of unitary time-evolution}

So far, we have considered mainly steady-state regime of
generation, including the effects of dissipation and
cavity-induced feedback. However, in order to better understand
the pecularities of the entanglement for the system under
consideration, it would be interesting and desirable to study the
situation, when the dissipation in the cavity is unessential and
the evolution of the modes, due to their interactions, is
described approximately by the master equation (\ref{MasterEq})
without losses. In our analysis we shall assume that both
subharmonic modes are initially in the vacuum state. Then, the
system's state in the absence of losses is generated from the
vacuum state by the unitary transformation
\begin{equation}
\left| \Psi \left( t\right) \right\rangle =u\left( t\right) \left|
0\right\rangle _{1}\left| 0\right\rangle _{2}=e^{-\frac{i}{\hbar }%
H_{int}t}\left| 0\right\rangle _{1}\left| 0\right\rangle _{2},
\label{psi}
\end{equation}
where
\begin{eqnarray}
H_{int}=i\hbar\frac{k
E}{\gamma_{3}}\left(a_{1}^{+}a_{2}^{+}-a_{1}a_{2}\right)
+\hbar\chi\left(a_{1}^{+}a_{2}-a_{1}a_{2}^{+}\right).
\label{UnitaryH}
\end{eqnarray}
Here we neglect the pump depletion effects considering classically
the amplitude of driving field as a constant and also consider the
case of zero-detunings $\Delta _{1}=\Delta _{2}=\Delta =0$. For
the cavity configuration considered the validity of such
approximation is guaranteed for short interaction time intervals
$(kE/\gamma_{3})^{-1},\chi^{-1}\lesssim \Delta t \ll \gamma^{-1}$,
provided that the coupling constants $kE/\gamma_{3}$ and $\chi$
exceed the dumping rates of the modes.

Note, that for $\chi\rightarrow 0$ the state (\ref{psi}) is
transformed to the well-known two mode squeezed state
\cite{Schumacher}. We note, that while the entangled two-mode
squeezed vacuum state is achieved in the laboratory (in the
nondegenerate parametric amplifier \cite{levon2}), the generation
of the state (\ref{psi}) is an interesting option for the future.
From an experimental point of view it can be realized in the
combined scheme, that is parametric down conversion and wave
plate, driven by a pulsed pump field. It is obvious, that the
interaction time $\Delta t$ in this case is limited by the
duration $T$ of laser pulses, $\Delta t \lesssim T$. Roughly
speaking, such scheme is similar to the optical nondegenerate
parametric amplifier in the presence of phase-locking process,
where the interaction time is increased by a cavity for both
subharmonic modes. However, we note again, that a complete study
of the experimental configuration related to the nondissipative
case is beyond the scope of this paper.

\begin{figure}
\includegraphics[angle=-90,width=0.48\textwidth]{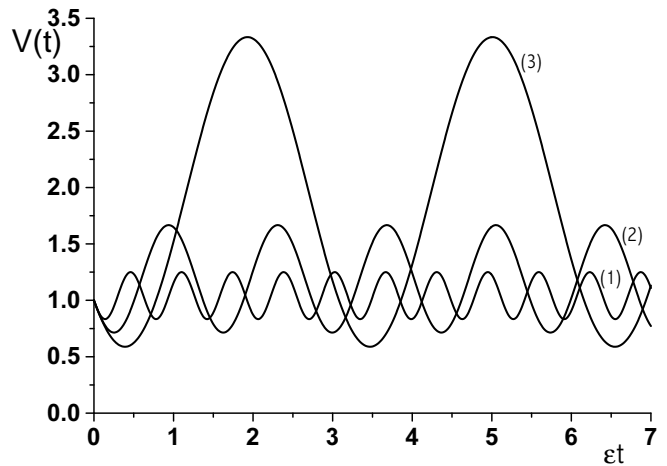}
\caption{ Unitary evolution of the variance $V\left(t\right)$
versus the scaled dimensionless interaction time $\varepsilon t$
for the range of $\varepsilon <\chi $ and provided that $\cos
\left( \Sigma\theta\right) =1$. The parameters are: $\varepsilon
/\chi =0.2$ (curve 1), $\varepsilon /\chi =0.4$\ (curve 2) and
$\varepsilon /\chi =0.7$ (curve 3). } \label{cos_fig}
\end{figure}

Now let us examine the entangled properties of the resultant state
(\ref{psi}). Using formula (\ref{Vby_a}), where $n=\langle
u^{-1}(t)a_{i}^{+}a_{i}u(t)\rangle $ and $\left\langle
a_{1}(t)a_{2}(t)\right\rangle =\langle
u^{-1}(t)a_{1}a_{2}u(t)\rangle$, after a long but straightforward
algebra we get the final result for the variance
$V=\frac{1}{2}(V_{+}+V_{-})$ for two different operational
regimes: $\varepsilon <$ $\chi $ and $\varepsilon >$ $\chi $. The
variance for the range $\varepsilon <$ $\chi $ of a comparatively
weak scaled pump field $\varepsilon =kE/\gamma _{3}$ reads as
\begin{eqnarray}
V\left(t\right)&=&1-\frac{\varepsilon ^{2}}{\mu ^{2}}\left[\cos (2\mu t)-1\right]  \nonumber \\
&-&\frac{\varepsilon }{\mu }\sin (2\mu t)\cos \left(\Sigma\theta
\right) , \label{VShrodingerBelow}
\end{eqnarray}
where $\mu $ $=\sqrt{\chi ^{2}-\varepsilon ^{2}}$. Thus, we
observe a periodic evolution of $V$ in this regime typical for the
linear coupling. The level of squeezing of the two modes is
periodically repeated. The behavior of the two-mode variance also
significantly depends on the phase matching condition, so that we
may tune the phase sum to maximize the entanglement. We further
choose for illustrations the phase sum corresponding to $\cos
\left(\Sigma\theta\right) =1$ for both operational regimes. It
seems perhaps more intuitive to study the time-behavior scaled on
the parameter $\varepsilon =kE/\gamma _{3}$, which contains the
pump field amplitude and is an adjustable parameter of the system.
The dependence of $V$ versus the scaled time-interval is shown in
Fig.~\ref{cos_fig}, where the three curves correspond to three
different choices of the ratio $\varepsilon /\chi$. Common to all
curves is that the variance is nonclassical and squeezed at least
at the points of its minima $t_{\min }$, which can be obtained by
the formula $ctg\left(2\mu t_{\min }\right)=\varepsilon /\mu $. In
all cases, the maximal degree of two-mode squeezing
$V_{\min}=V\left(t\right)|_{t=t_{\min}}=0.5$ is achieved in the
limit $\varepsilon \rightarrow \chi $. For the curves on the
Fig.~\ref{cos_fig} the time-intervals corresponding to the minimal
values of $V_{\min}$ equal to $\varepsilon t_{\min
}\simeq{0.14}+{0.20}\pi k$(curve 1), $\varepsilon t_{\min
}\simeq{0.25}+{0.44}\pi k$(curve 2), $\varepsilon t_{\min
}\simeq{0.39}+{0.98}\pi k$(curve 3), ($k=0,1,2,...$).

If the opposite inequality holds, $\varepsilon >$ $\chi $, then
the nonlinear parametric interaction becomes dominant over the
linear coupling and the variance is given by the following formula
\begin{eqnarray}
V\left(t\right)&=&1+\frac{\varepsilon ^{2}}{\eta ^{2}}\left[\cosh \left(2\eta t\right)-1\right]  \nonumber \\
&-&\frac{\varepsilon }{\eta }\sinh\left(2\eta t\right)\cos \left(
\Sigma\theta\right) , \label{VShrodingerAbove}
\end{eqnarray}
where $\eta =\sqrt{\varepsilon ^{2}-\chi ^{2}}$.

\begin{figure}
\includegraphics[angle=-90,width=0.48\textwidth]{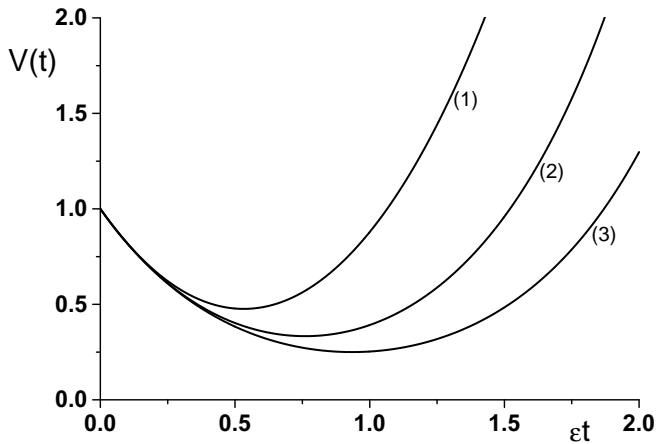}
\caption{ Unitary time evolution of the variance
$V\left(t\right)\;$ for the range of $\varepsilon >\chi$ and
provided that $\cos \left( \Sigma\theta\right) =1$. The parameters
are: $\varepsilon /\chi =1.1$ (curve 1), $\varepsilon /\chi =2$\
(curve 2) and $\varepsilon /\chi =3$ (curve 3).} \label{exp_fig}
\end{figure}

For $\chi\rightarrow 0$ and $\cos \left( \Sigma\theta\right) =1$,
from this formula we arrive at a well-known result,
$V\left(t\right)=\exp \left( -2\varepsilon t\right) $, for
two-mode squeezed state. This shows that in the limit of infinite
squeezing the corresponding state approaches to a simultaneous
eigenstate of $X_{1}-X_{2}$ and $Y_{1}+Y_{2}$ , and thus becomes
equivalent to the EPR state. Fig.~\ref{exp_fig} shows the behavior
of $V$ when the system operates in the regime $\varepsilon >$
$\chi $. This figure clearly shows that as the interaction time
increases, the variance decreases and reaches its minimum. Then
the squeeze variance exponentially increases with the growth of
the interaction time. For the data in Fig.~\ref{exp_fig} we obtain
that the time intervals for which the variance reaches the minima
are $\varepsilon t_{\min }\simeq{0.53}$(curve 1), $\varepsilon
t_{\min }\simeq{0.76}$ (curve 2), $\ \varepsilon t_{\min
}\simeq{0.93}$(curve 3). It is easy to check that these points of
minima can be found by the formula $ctgh(2\eta t_{\min
})=\varepsilon /\eta$, provided that $\cos \left(
\Sigma\theta\right) =1$.

We also conclude that the variance squeezes up to a certain
interaction time, if $\varepsilon >$ $\chi $. It should be noted
that although the time evolution of the variances are quite
different for each of the operational regimes, the minimal values
of the variance are described by the formula which is the same for
both regimes
\begin{equation}
V_{\min}=V\left( t \right)|_{t = t _ {\min} }=\frac{\chi
}{\varepsilon +\chi }. \label{VminShrodinger}
\end{equation}

With increasing $\varepsilon /\chi $, in the regime $\varepsilon
>$ $\chi $, the minimal value of the variance decreases as
$V_{\min}\sim $ $\chi /\varepsilon \ll 1$, which means that
perfect squeezing takes place in the limit of infinite pump field.
This result is not at all trivial for the system considered, even
in the absence of dissipation and cavity-induced feedback, because
the insertion of polarization mixer usually destroys the two-mode
squeezing produced by nondegenerate parametric down-conversion.
The reason is that two-mode squeezed vacuum state is a
supperposition of two-photon Fock states $\left| n\right\rangle
_{1}\left| n\right\rangle _{2}$ and the polarization mixer
destroys the Fock states having the same number of photons, i.e.,
$n_{1}=n_{2}=n$. The detailed analysis of this problem can be
found, for example, in \cite{kim}.

Having discussed the CV entanglement for non-dissipative dynamics
we now turn our attention to the generation of entangled states of
light beams in self-phase locked NOPO completely taking into the
consideration dissipation and cavity-induced feedback. The results
will be presented in a steady-state regime of generation.

\subsection{Entanglement in the self-phase locked NOPO:
Sub-threshold regime}

Using formulas (\ref{dadbMeanFinal}), (\ref{VminBy_a}), after some
algebra, we obtain the minimal variance for the steady-state
regime in the following form
\begin{eqnarray}
V=1+\frac{\varepsilon \left( \varepsilon S^{2}-\sqrt{\gamma ^{2}
S^{4}+\Delta ^{2}\left(S^{2} - 2\chi ^{2}\right)^{2}}
\right)}{S^{4}-4\Delta^{2}\chi^{2}}. \label{VminBelow}
\end{eqnarray}

\begin{figure}
\includegraphics[angle=-90,width=0.48\textwidth]{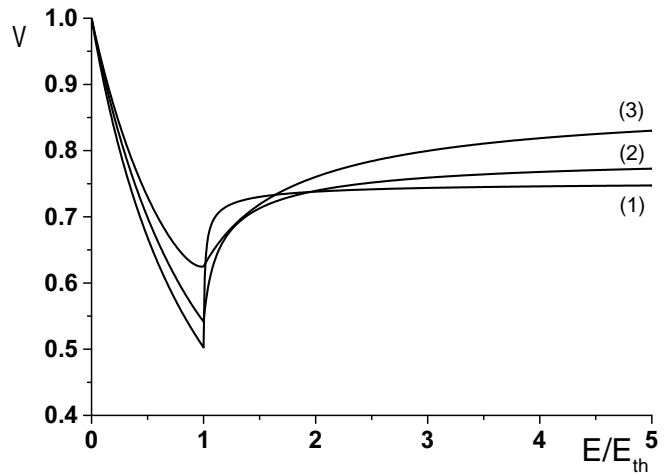}
\caption{The minimized variance $V$ versus the dimensionless
amplitude of the pump field $E/E_{th}=\varepsilon /\varepsilon
_{th}$ for both operational regimes. $\varepsilon_{th}$ and
$E_{th}$ are given by the Eqs.(\ref{EpsilonThreshold}),
(\ref{EThreshold}). The parameters are: $\chi /\gamma =0.1$ (weak
coupling), $\Delta /\gamma =10$(curve 1); $\chi /\gamma
=0.5$(moderate coupling), $\Delta /\gamma =3$(curve 2) and $\chi
/\gamma =0.5$, $\Delta /\gamma =1$(curve 3).} \label{Vmin_fig}
\end{figure}

It is easy to check that in the limit $\chi \longrightarrow 0$\
the variance coincides with the analogous one for the ordinary
NOPO, $V=V_{-}=V_{+}=1-\varepsilon /(\varepsilon +\sqrt{\gamma
^{2}+\Delta ^{2}})$. We see that the minimal variance remains less
than unity for all values of pump intensity and is a monotonically
decreasing function of $\varepsilon ^{2}$. For all parameters the
maximal degree of two-mode squeezing $V\simeq 0,5$ is achieved
within the threshold range. It is also easy to check that this
expression is well-defined for all values of $\varepsilon ^{2}$,
including the vicinity of the threshold. One should keep in mind,
however, that the linear approach used does not describe the
threshold range where the level of quantum fluctuations increases.
As a consequence, some matrix elements of (\ref{dadbMeanFinal}),
(\ref{dn+dfi+MeanFinal}), (\ref{dn-dfi-MeanFinal}) increase
infinitely in the vicinity of threshold. Nevertheless, it follows
from (\ref{VminBelow}) and further results
(\ref{VminAbove})-(\ref{VMinusAbove}), that such infinite terms
are cancelled in the variance $V$, as well as in $V_{-},\;V_{+}$
for both operational regimes. This is not surprising, since such
cancellation of infinities in the quadrature amplitude variances
takes place for the ordinary NOPO also.

One of the differences between the squeezing effects of the
ordinary and self-phase locked NOPO is that the variances $V_{-}$
and $V_{+}$ for the ordinary NOPO are equal to each other, while
for the self-phase locked NOPO they are in general different. The
values of the non symmetric variances are expressed through $R$
according to formula (\ref{V+-ByVandR}). This quantity can be
calculated with the help of the formula (\ref{RBy_aForMin}). The
result is found to be

\begin{widetext}
\begin{equation}
R=\frac{\varepsilon \chi \Delta } {S^{4}-4\Delta^{2}\chi^{2}}
\left[ \frac { \varepsilon ^{4}-\left( \gamma ^{2}+\left(\Delta
-\chi \right)^{2}\right) \left( \gamma ^{2}+\left(\chi +\Delta
\right)^{2}\right) } { \sqrt{\gamma ^{2} S^{4}+\Delta
^{2}\left(S^{2} - 2\chi ^{2}\right)^{2} } } +2\varepsilon \right]
. \label{RForMinBelow}
\end{equation}
\end{widetext}

As we see, the final expressions (\ref{VminBelow}), (\ref
{RForMinBelow}) below the threshold are rather unwieldy. We show
the corresponding numerical results on the Figs.~\ref{Vmin_fig},
\ref{Vplusminus_fig}, \ref{VplusMulVminus_fig} for illustration.

\subsection{Entanglement in the self-phase locked NOPO:
Above-threshold regime}

Performing calculations for each of the cases $\Delta >0$ and
$\Delta <0$, we find the variance $V$ in the above-threshold
regime in the following form
\begin{equation}
V=\frac{3}{4}-\frac{1}{4\sqrt{1+\left(\varepsilon _{th}/\gamma
\right)^{2}\left( \left(\varepsilon/\varepsilon
_{th}\right)^{2}-1\right)}}+\frac{\chi }{4\mid \Delta \mid }.
\label{VminAbove}
\end{equation}
To rewrite this expression through the original parameters we
should take into account that $E/E_{th}=\varepsilon /\varepsilon
_{th}$ and
$\varepsilon_{th}^{2}/\gamma^{2}=\left(\chi/\gamma-\left| \Delta
\right|/\gamma\right)^{2}+1$. The results (\ref{VminBelow}) and
(\ref{VminAbove}) for both operational regimes are summarized in
Fig.~\ref{Vmin_fig}, where the variance $V$ is plotted as a
function of the amplitude of the pump field. One can immediately
grasp from the figure that the sum of variances
$V=\frac{1}{2}\left( V_{+}+V_{-}\right)$ remains less than unity
for all nonzero values of the pump field, provided that $\chi<
\left| \Delta \right|$. This shows the nonseparability of the
generated state. The maximal degree of entanglement is achieved in
the vicinity of threshold, $V\simeq 0,5$, if $\chi /\left| \Delta
\right| \ll 1$. In far above the threshold, $E>>E_{th}$, $V$
increases with mean photon numbers of the modes and reaches the
asymptotic value $V=3/4+\chi /4\left| \Delta \right|$. It should
also be mentioned that the result (\ref{VminAbove}) is expressed
through the scaled pump field amplitude $ \varepsilon =kE/\gamma
_{3}$ and hence depends on coupling constants $k$ and $\chi$.

We stress, that the level of two-mode squeezing in the proposed
scheme is limited due to the dissipation and pump depletion
effects and reaches only $50\%$ relative to the level of vacuum
fluctuations if the pump intensity draw near to the generation
threshold. Such limitation on the level of intracavity two-mode
squeezing takes place also for the ordinary NOPO as was been
demonstrated in Refs. \cite{new, Dechoum} by analysis of quantum
fluctuations in the near-threshold range. Analogous conclusion has
been made for one-mode squeezed light generated in OPO
\cite{Chaturvedi}. However, remind that these results pertain to
the full squeezing of intracavity variances and not the spectra of
the squeezing of output fields.

\begin{figure}
\includegraphics[angle=-90,width=0.48\textwidth]{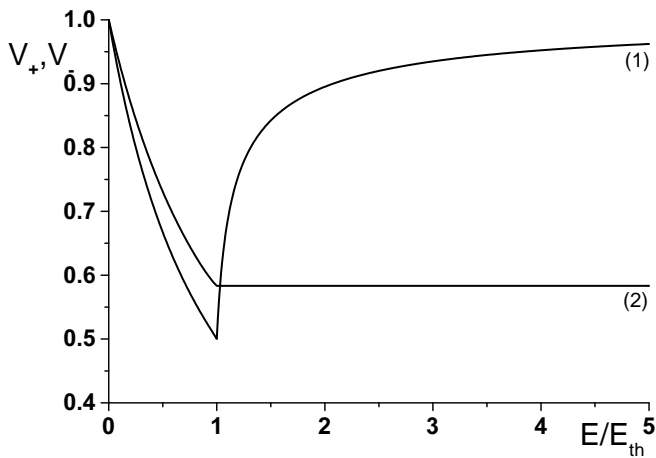}
\caption{ The variances $V_{+}\;$(curve 1) and $V_{-}\;$(curve 2)
versus the dimensionless amplitude of the pump field
$E/E_{th}=\varepsilon /\varepsilon _{th}$ for the parameters:
$\chi /\gamma =0.5,\;\Delta /\gamma =3$.} \label{Vplusminus_fig}
\end{figure}

As the last step in connecting two-mode squeezing to observables
of CV entanglement, we report the expressions of the non symmetric
variances calculated with the help of the formulas
(\ref{V+-ByVandR}), (\ref{VminBy_a} ) and (\ref{RBy_aForMin}).
Upon evaluating all required expectation values, we obtain
\begin{equation}
R=\frac{sgn(\Delta )}{4}\left(\frac{\mid \Delta \mid - \chi }{\mid
\Delta \mid }-\frac{1}{\sqrt{1+\left( \varepsilon ^{2}-\varepsilon
_{th}^{2}\right) /\gamma ^{2}}}\right) \label{RForMinAbove}
\end{equation}
and hence
\begin{eqnarray}
V_{+} &=&\frac{3\mid \Delta \mid+\chi+ sgn\left(\Delta \right)\cos
\left(\Delta\theta\right)\left(\mid \Delta \mid-\chi\right)}{4\mid \Delta \mid} \nonumber \\
&-&\frac{{1}+sgn\left(\Delta \right)\cos
\left(\Delta\theta\right)}{{4}\sqrt{1+\left( \varepsilon
^{2}-\varepsilon _{th}^{2}\right) /\gamma ^{2}}} ,
\label{VPlusAbove}
\end{eqnarray}
\begin{eqnarray}
V_{-} &=&\frac{3\mid \Delta \mid+\chi- sgn\left(\Delta \right)\cos
\left(\Delta\theta\right)\left(\mid \Delta \mid-\chi\right)}{4\mid \Delta \mid} \nonumber \\
&-&\frac{{1}-sgn\left(\Delta \right)\cos
\left(\Delta\theta\right)}{{4}\sqrt{1+\left( \varepsilon
^{2}-\varepsilon _{th}^{2}\right) /\gamma ^{2}}} ,
\label{VMinusAbove}
\end{eqnarray}
For the case of $\Delta\theta=0$, when the variances are maximally
different, the results are reduced to
\begin{eqnarray}
V_{-}&=&\frac{1}{2}+\frac{\chi }{2\Delta },\nonumber \\
V_{+}&=&1-\frac{1}{2\sqrt{1+\left(\varepsilon _{th}/\gamma
\right)^{2}\left( \left(\varepsilon/\varepsilon
_{th}\right)^{2}-1\right)}} \label{VPlusVMinusMaxDiff}
\end{eqnarray}
for $\Delta <0$. The case of $\Delta >0$ is obtained from (\ref
{VPlusVMinusMaxDiff}) by exchanging $V_{+}\rightarrow V_{-}$\ and $%
V_{-}\rightarrow V_{+}$. These results for both operational
regimes are depicted on Fig.~\ref{Vplusminus_fig}. Some features
are immediately evident. First of all one can see that both
variances are minimal in the critical range but show quite
different dependences on the ratio $\varepsilon /\varepsilon
_{th}$ above the threshold. Attentive reader may ask about
dependence of the results obtained on parametric coupling constant
$k$. We note in this connection that the squeezed variances are
expressed through the scaled pump field amplitude $\varepsilon
=kE/\gamma _{3}$ and hence depend in general on both coupling
constants $k$ and $\chi $.

\begin{figure}
\includegraphics[angle=-90,width=0.48\textwidth]{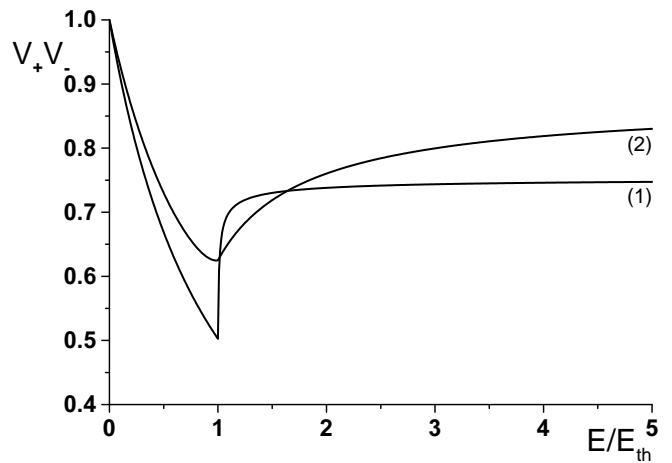}
\caption{ The product of the variances $V_{+}\;$ and $V_{-}\;$
versus the dimensionless amplitude of the pump field
$E/E_{th}=\varepsilon /\varepsilon _{th}$ for the parameters $\chi
/\gamma =0.1,\;\Delta /\gamma =10\;$(curve 1) and $\chi /\gamma
=0.5,\;\Delta /\gamma =1\;$(curve 2).} \label{VplusMulVminus_fig}
\end{figure}

In terms of demonstrating the CV strong EPR entanglement, one has
to apply another criterion $V_{+}V_{-}<\frac{1}{4}$. For the case
of the symmetric uncertainties ($V_{+}=V_{-}=V$), the product of
the variances $V^{2}\geq \frac{1}{4}$ and hence the strong EPR
entanglement can not be realized. In the general case, the product
of the variances reads as
\begin{equation}
V_{+}V_{-}=V^{2}-R^{2}\cos ^{2}\left(\Delta\theta\right) .
\label{V+V-AboveGeneral}
\end{equation}
It seems that $V_{+}V_{-}$ lies below $1/4$ at least for the
relative phase $\Delta\theta=\pm \pi m$, $(m=1,2,...)$, and in the
vicinity of threshold, where $V\simeq 0,5$. However, for such
selection of the phases we arrive at
\begin{equation}
V_{+}V_{-}=\frac{\mid \Delta \mid + \chi}{4\mid \Delta \mid}
\left( 2- \frac{1}{\sqrt{1+\left(\varepsilon _{th}/\gamma
\right)^{2}\left( \left(\varepsilon/\varepsilon
_{th}\right)^{2}-1\right)}}\right). \label{V+V-AboveFinal}
\end{equation}
It is easy to check that the product of the variances exceeds
$\frac{1}{4}$ even in the vicinity of the threshold. This quantity
for both operational regimes is illustrated in
Fig.~\ref{VplusMulVminus_fig}. It should be noted again that a
detailed analysis of this problem, must include more accurate
consideration of quantum fluctuations in the critical ranges.

\section{CONCLUSION}

Our work demonstrates the possibility of creation of CV entangled
states of light beams in the presence of phase localizing
processes. We have called such quantum states the entangled
self-phase locked states of light \cite{ourPaper} and we have
shown that they can be generated in NOPO recently realized in the
experiment \cite{mason}. This device is based on the type-II
phase-matched down-conversion and additional phase localizing
mechanisms stipulated by the intracavity waveplate. The novelty is
that this device provides high level of phase coherence between
the subharmonics in contrast to what happens in the case of
ordinary NOPO, where the phase diffussion takes place. We have
shown that both two mode squeezing and quantum phase locking
phenomena are combined in such NOPO. This development paves the
way towards the generation of bright CV entangled light beams with
well-localized phases. It looks like that this scheme involving
phase locking may be potentially useful for precise
interferometric measurements and quantum communications, because
it combines quantum entanglement and stability of type - II phase
matching with effective suppression of phase noise. Particularly,
in practice, this phase locked scheme can effectively be used in
homodyne detection setups, as well as for observation of an
interference with nonclassical light beams. We believe, that such
source of bright entangled light providing phase coherence will
also be applicable for realizing CV quantum teleportation. The
price one has to pay for these advantages is the small aggravation
of the degree of CV entanglement in self-phase locked NOPO in
comparison with the case of ordinary NOPO. The quantum theory of
self-phase locked NOPO has been developed in linear treatment of
quantum fluctuations for both below- and above-threshold regimes
of generation. We have studied the CV entanglement as two-mode
squeezing and have shown that entanglement is present in the
entire range of pump intensities. In all cases the maximal degree
of two-mode squeezing $V\simeq 0,5$ is achieved in the vicinity of
the threshold, provided that the coupling constant $\chi$ is much
less than the mode detunings. We have demonstrated that, as a
rule, highest degree of CV entanglement occurs for weak linear
coupling strength $\chi$. It has also been shown that the amount
of the entanglement can be controlled via the phase difference
$\Phi_{\chi}$. The other peculiarities of the system of interest
have been established for the case of unitary dynamics which may
be realized for the short interaction times in the case of pulsed
pump fields. One of these concerns the presence of two operational
regimes generating two-mode squeezing. If the linear coupling
between subharmonics dominates over the parametric
down-conversion, $\varepsilon<\chi$, we have observed a periodic
evolution of the squeezed variance. Maximal degree of squeezing
has been $V\simeq 0,5$ in this regime. If the parametric
interaction becomes dominant, $\varepsilon>\chi$, the more high
degree of two-mode squeezing can be obtained,
$V_{\min}=\frac{\chi}{\varepsilon+\chi}<{0.5}$, but up to certain
interaction time.

In our analysis we have not investigated all possible quantum
effects of self-phase locked parametric dynamics. In particular,
we have noted that the system considered displays different types
of quantum correlations, but we have not analyzed their connection
with all possible kinds of the entanglement. This analysis might
involve also the case of unitary dynamics, where the generalized
two-mode squeezed state has been presented. Consideration of
quantum fluctuations in the near threshold operational range of
self-phase locked NOPO also deserves special attention for more
accurate identification of CV entanglement. These topics are
currently being explored and will be the subject of forthcoming
work.

\begin{acknowledgments}
Acknowledgments: We thank J. Bergou for helpful discussions. This
work was supported by the NFSAT PH 098-02/CRDF 12052 grant and
ISTC grant no. A-823.
\end{acknowledgments}

\end{document}